\documentclass{article}

\usepackage{amssymb,amsfonts,amsmath,stmaryrd}
\usepackage{cite,enumerate,float,indentfirst}
\usepackage{color}

\def\be{\begin{eqnarray}}
\def\ee{\end{eqnarray}}
\def\nn{\nonumber}

\definecolor{red}{rgb}{1,0,0}
\definecolor{orange}{rgb}{1,0.5,0}
\definecolor{violet}{rgb}{0.7,0,1}

\hoffset= - 1.0in         

\begin{document}

\hfill ITEP/TH-02/19

\hfill IITP/TH-02/19

\hfill MIPT/TH-02/19

\bigskip

\bigskip

\bigskip

\bigskip

\centerline{\Large{
On exclusive Racah matrices $\bar S$ for rectangular representations
}}

\bigskip

\centerline{\bf A.Morozov}

\bigskip

\centerline{\it ITEP, IITP \& MIPT, Moscow, Russia}

\bigskip

\bigskip

\bigskip

\bigskip

\centerline{ABSTRACT}

\bigskip

{\footnotesize
We elaborate on the recent observation that evolution
for twist knots simplifies when described in terms of
triangular evolution {\it matrix} ${\cal B}$,
not just its eigenvalues $\Lambda$,
and provide a {\it universal} formula for ${\cal B}$,
applicable to arbitrary rectangular representation $R=[r^s]$.
This expression is in terms of skew characters and
it remains literally the same for the 4-graded
rectangularly-colored hyperpolynomials,
if characters are substituted by Macdonald polynomials.
Due to additional factorization property of the differential-expansion
coefficients for the double-braid knots,
explicit knowledge of twist-family evolution leads
to a nearly explicit answer for
Racah matrix $\bar S$ in arbitrary rectangular representation $R$.
We also relate matrix evolution to existence of a peculiar rotation $U$
of Racah matrix, which diagonalizes the $Z$-factors in the differential
expansion -- what can be a key to further generalization to non-rectangular
representations $R$.

}

\bigskip

\bigskip

\bigskip

\bigskip

\bigskip

\bigskip

Knot polynomials \cite{knotpols} belong to an advanced chapter of modern mathematical
physics.
Their understanding would provide a non-trivial extension of our knowledge
from two to three dimensions and is foreseeably important for various branches
of science.
One of the intimately related is the theory of Racah matrices \cite{Racah}
which was supposed to help in the study of knot polynomials.
Somewhat surprisingly, however, the inspiration went in the opposite direction --
it turns simpler to find the knot polynomials and then convert them into
the new results for Racah matrices.
This paper describes a spectacular result of this kind.

\bigskip

\bigskip

\bigskip

{\bf 1.} Conceptually, knot invariants are the Wilson loop averages
\be
{\cal H}_R^{\cal C}(q,A) = \left<{\rm Tr}_R\, P\exp\oint_{\cal C}{\cal A} \right>
\ee
i.e. the central objects in the Yang-Mills theory,
but they are evaluated in one of its simplest versions --
the topological $3d$ Chern-Simons theory \cite{CS}.
This reduces the most interesting dependence on the integration contour ${\cal C}$
to that on its topological class (knotting).
Still, even this dependence is highly non-trivial --
nothing to say about that on the coupling constant $g$,
on the rank of gauge group ${Sl}(N)$
and on its representation.
One of the spectacular facts is that the average can be calculated exactly,
and the it turns to be a polynomial of peculiar
non-perturbative variables $q=\exp\left(\frac{2\pi i}{g+N}\right)$ and $A=q^N$.
The reason which makes the theory solvable is that it describes the peculiar
time-evolution of $2d$ conformal blocks \cite{CFT}, defined by their monodromies.
This makes the study of knot polynomials the next step after a deep understanding
of conformal blocks, achieved recently through the theory of Nekrasov functions \cite{Nek}
(instanton calculus in $4d$ and $5d$ supersymmetric Yang-Mills theories \cite{LMNS})
and the AGT relations \cite{AGT}.
On this way one can expect new insights and discoveries --
and they are indeed being intensively produced and shed new light on a variety
of subjects: from non-perturbative calculations and associated extension of
integrability theory to the hard problems in conventional representation theory.

\newpage

One of such problems is calculation of Racah matrices ($6j$-symbols) --
intensively studied in theoretical physics and included into the standard
textbooks like \cite{LL3}.
Conformal block is a contribution of particular intermediate state  into the correlator,
e.g. in the 4-point case

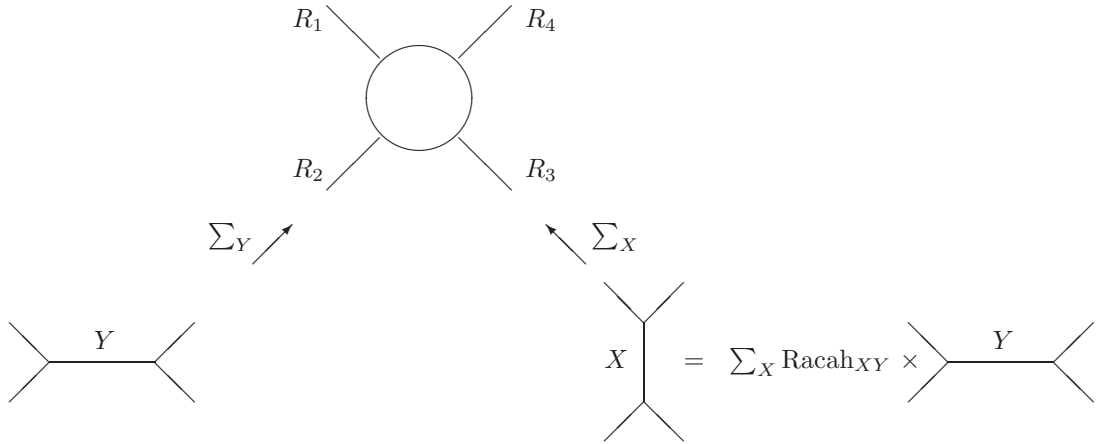
\begin{picture}(300,200)(-200,-140)
\put(0,0){\circle{40}}
\put(15,15){\line(1,1){20}}\put(15,-15){\line(1,-1){20}}
\put(-15,15){\line(-1,1){20}}\put(-15,-15){\line(-1,-1){20}}
\put(-48,-30){\mbox{$R_2$}}
\put(-48,27){\mbox{$R_1$}}
\put(40,-30){\mbox{$R_3$}}
\put(40,27){\mbox{$R_4$}}
\put(-63,-63){\vector(1,1){15}}
\put(-80,-55){\mbox{$\sum_Y$}}
\put(63,-63){\vector(-1,1){15}}
\put(65,-55){\mbox{$\sum_X$}}
\put(-140,-100){
\put(0,0){\line(1,0){40}} \put(17,5){\mbox{$Y$}}
\put(0,0){\line(-1,1){15}}\put(0,0){\line(-1,-1){15}}
\put(40,0){\line(1,1){15}}\put(40,0){\line(1,-1){15}}
}

\put(85,-100){
\put(0,-15){\line(0,1){30}} \put(-15,-2){\mbox{$X$}}
\put(0,-15){\line(-1,-1){15}}\put(0,15){\line(-1,1){15}}
\put(0,15){\line(1,1){15}}\put(0,-15){\line(1,-1){15}}
\put(115,0){
\put(-100,-2){\mbox{$=\ \ \sum_X {\rm Racah}_{XY}\ \times $}}
\put(0,0){\line(1,0){40}} \put(17,5){\mbox{$Y$}}
\put(0,0){\line(-1,1){15}}\put(0,0){\line(-1,-1){15}}
\put(40,0){\line(1,1){15}}\put(40,0){\line(1,-1){15}}
}}

\end{picture}

\noindent
where two different ways of decomposition are shown,
and correlator is the sum over $Y$ or, alternatively, over $X$.
The blocks in the $t$-channel are related to those in the $S$-channel
by a linear transformation -- and it is defined by the Racah matrix.
In fact, when external lines belong to irreducible representations of the
symmetry group, Racah matrix depends only on representations,
not on the choice of particular elements and can be considered
as defining the associativity relation in representation products:
\be
{\rm Racah}:  \ \ \ \ \
\Big(
(R_1\otimes R_2) \otimes R_3 \longrightarrow R_4\Big)
\ \ \ \longrightarrow \ \ \
\Big(R_1\otimes (R_2\otimes R_3) \longrightarrow R_4\Big)
\ee
It is a matrix w.r.t. to the intermediate representations $X\in R_1\otimes R_2$
and $Y\in R_2\otimes R_3$.
Clearly Racah matrices are crucial for all kinds of dualities in string theory
and their study for various kinds of groups and representations
remains one of the central problems of modern theory.
Unfortunately, this is a very hard calculational problem and Racah matrices
are not actually known even for the finite dimensional irreducible representations
of $Sl_N$ group, labeled by Young diagrams -- nothing to say about the more
important example of DIM algebras, with representations, labeled by plane partitions.

Calculation of colored (i.e. with non-trivial diagrams $R$) Wilson averages
in Chern-Simons theory is reduced to convolutions of Racah matrices by the
modern version of Reshetikhin-Turaev formalism \cite{RT} --
however, until recently these matrices were actually unknown for Young diagrams $R$ which
have more than a single column or line.
In \cite{M16} it was suggested to reverse the logic -- and extract Racah matrices
from non-trivial properties of knot polynomials for some relatively simple
knot families, like their evolution and differential expansion properties.
This appeared a very fruitful idea, and it led to explicit evaluation of Racah
matrices for all 2-line representations $R$, some of which are presented
in \cite{knotebook}.
However, originally it was a hard task, and progress depended on the understanding
of the structure of the formulas, which has been slowly revealed in a sequence of works.
A new progress was achieved in a recent ref.\cite{KNTZ} (KNTZ),
which, as we are going to demonstrate
in the present paper, leads to a great simplification and provides explicit answers
for "exclusive" Racah matrices $\bar S$ (see below) in arbitrary rectangular
representations $R$: as we explain, they have the form of finite sums
\be
{\bar S}^R_{\mu\nu} \sim \sum_{\lambda\in R}
Z_R^\lambda\cdot {\cal E}_{\lambda\mu}{\cal E}_{\lambda\nu}
\ee
over sub-diagrams $\lambda$ of $R$ with explicitly known factors $Z$ and
matrices ${\cal E}$, which are the eigenfunction (diagonalization) matrices
of {\it triangular} ${\cal B}$, explicitly expressed through the well-known
skew Schur functions.
This is a tremendously simple expression, which can hardly be further simplified:
in general case there are no more cancellations between the terms
of the remaining sum.

Though exhaustive and explicit, this expression is just a  conjecture,
depending on a number of mysterious observations made in \cite{M16} and afterwards.
It passed a great number of highly non-trivial cross-checks -- beginning from unitarity of
the so constructed matrix $\bar S$ -- and there is a little doubt that the answer
is correct.
However, its derivation remains a big challenge for quantum field theory
and, most probably, will continue to be a source of new inspiration --
as it happened at the previous stages.

\bigskip
\newpage

{\bf 2.} As already stated,
the recent breakthrough in \cite{KNTZ} almost completed the quest for description
of rectangularly-colored knot polynomials for the double twist (double braid) knots,
originated in \cite{IMMMfe,evo} and \cite{M1605,M16},
and advanced in \cite{KM17fe}-\!\cite{M333}.
The crucial observation of \cite{M16} was that the coefficients of the differential
expansion for the double twist colored HOMFLY-PT polynomial \cite{knotpols}
factorize into a product of those for
the twist knots with $n=1$:


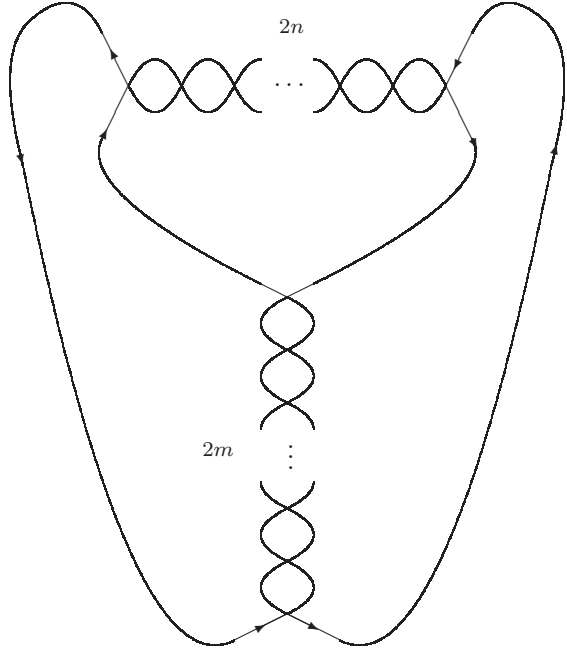
\begin{picture}(200,280)(-220,-230)
\qbezier(-40,0)(-50,20)(-60,0)
\qbezier(-40,0)(-50,-20)(-60,0)
\qbezier(-20,0)(-30,20)(-40,0)
\qbezier(-20,0)(-30,-20)(-40,0)
\qbezier(-20,0)(-15,10)(-10,10)
\qbezier(-20,0)(-15,-10)(-10,-10)
\put(-5,0){\mbox{$\ldots$}}
\qbezier(10,10)(15,10)(20,0)
\qbezier(10,-10)(15,-10)(20,0)
\qbezier(20,0)(30,20)(40,0)
\qbezier(20,0)(30,-20)(40,0)
\qbezier(40,0)(50,20)(60,0)
\qbezier(40,0)(50,-20)(60,0)
\put(-60,0){\line(-1,2){10}}
\put(-60,0){\line(-1,-2){10}}
\put(60,0){\line(1,2){10}}
\put(60,0){\line(1,-2){10}}
\qbezier(0,-80)(-20,-90)(0,-100)
\qbezier(0,-80)(20,-90)(0,-100)
\qbezier(0,-100)(-20,-110)(0,-120)
\qbezier(0,-100)(20,-110)(0,-120)
\qbezier(0,-120)(-10,-125)(-10,-130)
\qbezier(0,-120)(10,-125)(10,-130)
\put(0,-145){\mbox{$\vdots$}}
\qbezier(0,-160)(-10,-155)(-10,-150)
\qbezier(0,-160)(10,-155)(10,-150)
\qbezier(0,-160)(-20,-170)(0,-180)
\qbezier(0,-160)(20,-170)(0,-180)
\qbezier(0,-180)(-20,-190)(0,-200)
\qbezier(0,-180)(20,-190)(0,-200)
\put(0,-80){\line(-2,1){10}}
\put(0,-80){\line(2,1){10}}
\put(0,-200){\line(-2,-1){10}}
\put(0,-200){\line(2,-1){10}}
\put(0,-200){\line(-2,-1){20}}
\put(0,-200){\line(2,-1){20}}
\qbezier(-10,-75)(-80,-40)(-70,-20)
\qbezier(10,-75)(80,-40)(70,-20)
\put(-10,-205){\vector(2,1){2}}
\put(10,-205){\vector(2,-1){2}}
\put(-65,10){\vector(-1,2){2}}
\put(65,10){\vector(-1,-2){2}}
\put(-70,-20){\vector(1,2){2}}
\put(70,-20){\vector(1,-2){2}}
\put(-3,20){\mbox{\footnotesize$2n$}}
\put(-32,-140){\mbox{\footnotesize $2m$}}
%
\qbezier(-70,20)(-80,40)(-97,25)
\qbezier(-97,25)(-111,13))(-100,-30)
\qbezier(-100,-30)(-60,-230)(-20,-210)
\qbezier(70,20)(80,40)(97,25)
\qbezier(97,25)(111,13))(100,-30)
\qbezier(100,-30)(60,-230)(20,-210)
\put(-102,-22){\vector(1,-4){2}}
\put(100,-30){\vector(1,4){2}}
\end{picture}

\vspace{-0.7cm}

\be
\boxed{ \begin{array}{ccc}
{\cal H}_{R}^{(m,n)}
= \sum_{\lambda\subset R}
Z^\lambda_{R} \cdot {\cal F}_\lambda^{(m,n)}
= \sum_{\lambda\subset R}
Z^\lambda_{R} \cdot {F_\lambda^{(m)}F_\lambda^{(n)}}
\\ \\
F_\lambda^{(m)} =     \sum_{\mu\subset \lambda} f_{\lambda\mu}\cdot\Lambda_\mu^m
\end{array} }
\label{difexpan}
\ee

\bigskip

\noindent
The sum here goes over all the sub-diagrams $\lambda$
of the rectangular Young diagram $R=[r^s]$,
and the evolution eigenvalues $\Lambda_\mu$
are expressed through the hook parameters of the diagram:
\be
\mu=(a_1,b_1|a_2,b_2|\ldots) \ \ \  \Longrightarrow \ \ \
\Lambda_\mu
= \prod_h  (Aq^{a_h-b_h})^{2(a_h+b_h+1)}
= \!\! \prod_{h=1}^{\#\ {\rm of\ hooks}}\Big(A^2q^{2({\rm leg}_h-{\rm arm}_h)}\Big)^{\!\!\!
\overbrace{{\rm leg}_h+{\rm arm}_h+1}^{{\rm  length\ of\ the\ h-th\ hook}}}
\label{Lambda}
\ee
The weights in the sum come from the differential expansion  \cite{M1605,KM17fe}
for the figure-eight knot with $(m,n)=(1,-1)$
\be
Z_{[r^s]}^\lambda
= \left(-\frac{ (q-q^{-1})^2}{A^2} \right)^{|\lambda|}\cdot
\chi^*_{\lambda^{tr}}(r)\chi^*_\lambda(s)\cdot
\frac{\chi^*_\lambda(N+r)\chi^*_\lambda(N-s)}{(\chi_\lambda^\circ)^2}
\label{Zfactor}
\ee
Here $\chi^*_\lambda(r) := \chi_\lambda\{p^*(r)\}$ here and
$\chi^\circ_\lambda := \chi_\lambda\{p^\circ\}$, which we will also need in (\ref{Bels})
below,
denote restriction of the Schur functions to the special loci:
with time-variables $p_k$ substituted respectively by
\be
p_k^*(r):=\frac{q^{kr}-q^{-kr}}{q^k-q^{-k}}
\ \ \ \ \ \  {\rm and} \ \ \ \ \ \
p_k^\circ:= \frac{(q-q^{-1})^k}{q^k-q^{-k}}
\ee
Inverse of $\chi^\circ_\lambda$ differs by a power of $q$
from the $N$-independent denominator of $\chi_\lambda^*(N)$,
and (\ref{Zfactor}) combines the usual product of differentials, combinatorial factor and
$F^{(1)}_\lambda$.
The  arguments  of the HOMFLY-PT polynomial are $q$ and $A=q^N$.
Dependence on parameters $m$ and $n$ of the knot appears in (\ref{difexpan})
only through the powers of $\Lambda$,
while dependence on parameters $r$ and $s$ of representation --
only through the {\it knot-independent} weights $Z_{[r^s]}^\lambda$ in the sum,
which restrict the summation domain to  $\lambda\subset R=[r^s]$.
Note that in this paper we absorbed the combinatorial coefficients into $Z$-factors.

The analogues $C_{\mu\nu}$ of the Adams coefficients in the double-twist analogue of the
Rosso-Jones formula \cite{DMMSS,RJ}
\be
{\cal H}_{r^s}^{(m,n)} = \sum_{\mu,\nu\subset [r^s]} C_{\mu\nu}\Lambda_\mu^m\Lambda_\nu^n
\label{RJan}
\ee
\vspace{-0.3cm}
are therefore equal to
\vspace{-0.3cm}
\be
C_{\mu\nu} = \sum_{\mu,\nu\subset\lambda\subset [r^s]}
{Z_{[r^s]}^\lambda f_{\lambda\mu}f_{\lambda\nu}}
\label{Adan}
\ee
For the figure eight knot, unknot and the trefoil the $F$-functions are nearly trivial:
\be
F_\lambda^{(-1)}=1, \ \ \ F_\lambda^{(0)}=0, \ \ \
F_\lambda^{(1)}
= (-A)^{|\lambda|}\Lambda_\lambda^{1/2}
= \prod_{i=1}^{\#\ {\rm of\ hooks}} (-A^2q^{a_i-b_i})^{(a_i+b_i+1)} := \Lambda'_\lambda
\label{Ftref}
\ee
(the difference between the trefoil $F$-factor $\Lambda'$ and
the $\bar T^2$ eigenvalue $\Lambda$ from (\ref{Lambda})
is  a factor $2$ in the power of $q$).

\bigskip

{\bf 3.} Arborescent calculus \cite{arbor,arborgauge}
implies that the same HOMFLY polynomial
is expressed through Racah matrix $\bar S$:\ \
$
\Big((R\otimes \bar R)\otimes R \longrightarrow R \Big)
\ \stackrel{\bar S}{\longrightarrow} \
\Big(R\otimes (\bar R \otimes R) \longrightarrow R \Big)
$
by almost the same formula:
\be
{\cal H}_{R}^{(m,n)}
=  \sum_{\mu,\nu\subset R}
\frac{\sqrt{{\cal D}_\mu{\cal D}_\nu}}{\chi^*_R(N)}\,\bar S_{\mu\nu}^{R}\,
\Lambda_\mu^m\Lambda_\nu^n
\label{HRdb}
\ee
where for rectangular $R=[r^s]$ the ${\cal D}_\mu = \chi_R^*(N)^2\cdot C_{\emptyset \mu}
= \chi_{(\mu,\mu)}^*(N)$ are dimensions of the composite representations
$(\mu,\mu)\in R\otimes\bar R$, see eq.(28) in \cite{MMhopf} for an explicit expression.
From (\ref{HRdb}) we obtain:
\be
\boxed{
\bar S^{[r^s]}_{\mu\nu} = \frac{\chi^*_{[r^s]}}{\sqrt{{\cal D}_\mu{\cal D}_\nu}}
\sum_{\mu,\nu\subset\lambda\subset [r^s]}
Z_\lambda^{[r^s]} f_{\lambda\mu}f_{\lambda\nu}
}
\label{bS}
\ee
while another Racah matrix $S$:\ \
$
\Big((\bar R\otimes  R)\otimes R \longrightarrow R \Big)
\ \stackrel{\bar S}{\longrightarrow} \
\Big(\bar R\otimes (R \otimes R) \longrightarrow R \Big)
$
diagonalizes the product $\bar T\bar S\bar T$,
i.e. solves the linear equation
\vspace{-0.3cm}
\be
\sum_\mu  \bar T_\lambda\bar S_{\lambda\mu }\bar T_\mu S_{\mu\nu} T_\nu
= S_{\lambda\nu}
\label{SfrobS}
\ee
where $\bar T  $ and $T$ are known diagonal matrices, e.g.
$\bar T_{\mu\nu}^2 = \Lambda_\mu\cdot\delta_{\mu,\nu}$.
Orthogonality $\sum_\rho \bar S_{\mu\rho}\bar S_{\nu\rho} = \delta_{\mu\nu}$
of symmetric matrix $\bar S$ is equivalent to the sum rule
\be
\sum_{\rho\subset [r^s]} \frac{C_{\mu\rho}C_{\nu\rho}}{C_{\emptyset,\rho}} =
\chi^*_{[r^s]}(N)^2\cdot C_{\emptyset,\mu}\cdot\delta_{\mu,\nu}
\ee
for the combinations $C_{\mu\nu}$ in (\ref{Adan}), which does not contain
square roots $\sqrt{{\cal D}_\mu}$, what simplifies computer checks.

\bigskip

{\bf 4.}
For $F$ and $f$ ref.\!\cite{M16} suggested explicit expressions in a variety of examples,
which in \cite{KM17tw} were expressed in terms of skew characters.
However,  expression for generic $[r^s]$ was not found at that stage.
%
%
This was done in a recent paper \cite{KNTZ}.
After some polishing,
the observation there is that $F_\lambda^{(m)}$ actually have a very simple shape,
which one could (but did not) naturally anticipate from the evolution interpretation:
\be
\boxed{\boxed{
F_\lambda^{(m)} =   \sum_\mu \left({\cal B}^{m+1}\right)_{\lambda\mu}
}}
\label{evoF}
\ee
where ${\cal B}_{\lambda\mu}$ is a {\it triangular} matrix
with  the eigenvalues $\Lambda_\mu$ at diagonal and
non-vanishing entries only for embedded Young diagrams $\mu\subset\lambda$,
which are explicitly given by
\be
\boxed{
{\cal B}_{\lambda\mu} = (-)^{|\lambda|-|\mu|}\cdot \Lambda_\lambda\cdot
\frac{\chi_\mu^\circ \cdot \chi_{\lambda^\vee/\mu^\vee}^\circ}{\chi_\lambda^\circ}
}
\label{Bels}
\ee
$\vee$ stands for the transposition of Young diagrams.
$\chi_\lambda^\circ:=\chi_\lambda\{p^\circ\}$.
Like triangular $f$ in (\ref{difexpan}),
this matrix  ${\cal B}$ is an absolutely {\bf universal combinatorial quantity},
associated with embedding pattern of Young diagrams,
and completely independent of knots and particular rectangular representation $R$
(i.e. is independent of all the four parameters $n,m,r,s$).
The same is true about its eigenvector matrix ${\cal E}$ below.

\bigskip

{\bf 5.} As a simplest example,
the original result of \cite{evo} for symmetric representations $R=[r]$  is reproduced by
the following piece of ${\cal B}$, associated with the single-column diagrams $[r]$:
\be
{\cal B}_{ij} =\left\{
\begin{array}{c}
\overbrace{\frac{(-)^{i-j}}{q^{(i-2)(i-j)} }\cdot\frac{[i-1]!}{[j-1]![i-j]!}}^
{(-)^{i-j}\cdot\ \frac{\chi^\circ_{[j-1]}}{\chi^\circ_{[i-1]}}
\ \cdot\!\!\!\!\!\!\overbrace{\chi^\circ_{[1^{i-j}]}}^{\chi^\circ_{[i-1]^\vee/[j-1]^\vee}}}
\ \cdot\ \Lambda_{[i-1]} \\ \\
{\rm for}\ \ i\geq j \\ \\ \\
0   \\  {\rm for}\ \  i<j \\ \end{array}\right\}
= \ \left(\begin{array}{cccccc}
\Lambda_{[0]} & 0 & 0 & 0 & 0 & \ldots \\ \\
-\Lambda_{[1]} & \Lambda_{[1]} & 0 & 0 & 0 & \\ \\
\frac{1}{q^2}\Lambda_{[2]} & -\frac{[2]}{q}\Lambda_{[2]} & \Lambda_{[2]} & 0 & 0 &  \\ \\
-\frac{1}{q^6}\Lambda_{[3]} &\frac{[3]}{q^4}\Lambda_{[3]}
& -\frac{[3]}{q^2}\Lambda_{[3]} & \Lambda_{[3]} & 0 &
\\ \\
\frac{1}{q^{12}}\Lambda_{[4]} & -\frac{[4]}{q^9}\Lambda_{[4]}
&\frac{[4][3]}{[2]\,q^6}\Lambda_{[4]} & -\frac{[4]}{q^3}\Lambda_{[4]} & \Lambda_{[4]} & \\
\\
\ldots &&&&& \ldots
\end{array}\right)
\nn
\ee
Square brackets in this formula
are used to denote both the Young diagrams and quantum numbers
$[n]:=\frac{q^n-q^{-n}}{q-q^{-1}}$,
hopefully this does not cause a confusion.
Antisymmetric representations are controlled by a similar piece of ${\cal B}$,
associated with the sequence $[1^r]$.
In the case of $R=[3,3,3]$ the relevant fragment of ${\cal B}$ is $20\times 20$,
since there are $20$ sub-diagrams in $[3,3,3]$,
but the entries remain simple factorized expressions -- still this nicely reproduces
the complicated formulas for $F^{(m)}_\lambda$ from \cite{M333}.

\bigskip

{\bf 6.}
A possible way to explain the somewhat mysterious formula (\ref{evoF}) is to rewrite the
original arborescent formula for the HOMFLY-PT poly\-nomial
of the twist knot
\be
{\cal H}_R^{\text{twist}_m} =
D_R \cdot \Big(\bar S_R \bar T^{2} \bar S_R \bar T^{2m} \bar S_R\Big)_{\emptyset\emptyset}
\ee
with symmetric and orthogonal Racah matrix $\bar S_R$, $\bar S_R^2=I$, as
\be
H_R^{\text{twist}_m}
= D_R \cdot \Big(\bar S  \bar T^{2}    \
(\bar S  \bar T^{2} \bar S )^{m }\Big)_{\emptyset\emptyset}
= \sum_\lambda D_R \cdot
\Big(\bar S  \bar T^{2} \bar S  \bar T^{-2} \bar S \Big)_{\emptyset \lambda}
\Big((\bar S  \bar T^{2} \bar S )^{m+1}\Big)_{\lambda\emptyset}
\ee
and then further decompose it
by inserting the unity decomposition $I=U ^{-1}U$
with  an  auxiliary matrix $U$:
\be
H_R^{\text{twist}_m} = \sum_{\lambda }
\overbrace{D_R \cdot \Big(\bar S  \bar T^{2}
\underbrace{\bar S  \bar T^{-2} \bar S  U ^{-1}}_{U ^{-1}{\cal B}^{-1}}
\Big)_{\emptyset \lambda}}^{Z_R^\lambda} \cdot \overbrace{\sum_\mu
\Big(\boxed{\underbrace{U \bar S  \bar T^{2} \bar S  U ^{-1}}_{{\cal B}}}\Big)^{m+1}_{\lambda\mu}
\cdot U_{\mu\emptyset}}^{F_\lambda^{(m)}}
= \sum_{X} Z_R^\lambda \cdot
\overbrace{\sum_Y \left({\cal B}^{m+1}\right)_{\lambda\mu}}^{F_\lambda^{(m)}}
\label{arbortoKNTZ}
\ee
Thus we obtain {\it at once}, from a single $U$,
the decomposition formula (\ref{difexpan}) and the
matrix-evolution rule (\ref{evoF}).
However, we still need to choose $U$ appropriately, so that it provides
decomposition with the necessary (empirically justified) properties.
The last transition in (\ref{arbortoKNTZ}) requires that
$U$ has unities everywhere in the first column,
\be
U_{\mu\emptyset} =1
\label{UY0}
\ee
while its other elements are adjusted to make the KNTZ matrix
\be
\boxed{
{\cal B} := U\cdot \bar S  \bar T^{2} \bar S \cdot U ^{-1}
}
\label{BviaU}
\ee
{\it triangular} and satisfying the constraints
\be
\sum_{\mu} {\cal B}_{\lambda\mu} = \delta_{\lambda,\emptyset}
 \ \ \ \ \ \ \Longrightarrow \ \ \ \ \ \
\sum_Y ({\cal B}^2)_{\lambda\mu} =
 {\cal B}_{\lambda\emptyset}
\ \ \ \ \ \  \ \ \ \ \ \  \forall \lambda
\label{sumB}
\ee

\newpage

{\bf 7.}
Remarkably, after $U $ is adjusted to convert symmetric $\bar S \bar T \bar S $
into triangular ${\cal B}$, the matrix elements
\be
\boxed{
Z_R^\lambda :=  D_R \cdot
\Big(\bar S  \bar T^{2} \bar S  \bar T^{-2} \bar S \cdot U ^{-1} \Big)_{\emptyset \lambda}
}
\label{ZviaU}
\ee
appear to be factorized  for all {\bf rectangular representations} $R$
(for non-rectangular $R$ they are sums of several factorized expressions, see \cite{Mnonrect})
and reproduce the hook formulas for the Z-factors in
the differential expansions, in particular
\be
Z_R^\emptyset = D_R \cdot \Big(\bar S  \bar T^{2} \bar S  \bar T^{-2}
\bar S  U ^{-1} \Big)_{\emptyset \emptyset} = 1 \ \ \ \ \forall R
\ee
One more impressive fact is the factorization property,
which was the starting observation of \cite{M16}:
\be
H_R^{\text{double braid}_{m,n}} =
D_R \cdot \Big(\bar S  \bar T^{2n} \bar S  \bar T^{2m} \bar S \Big)_{\emptyset\emptyset}
= \ \sum_\lambda Z_R^\lambda \cdot \frac{F_\lambda^{(m)}\cdot F_\lambda^{(n)}}{F_\lambda^{(1)}}
\ee
It is now equivalent to a mysterious identity
\be
\Big(\bar S  \bar T^{2n} \bar S  \bar T^{2m} \bar S \Big)_{\emptyset\emptyset}
= \sum_\lambda
\overbrace{\Big(\bar S \bar T^{2} \bar S  \bar T^{-2} \bar S \, U ^{-1}
\Big)_{\emptyset \lambda}}^{Z^\lambda}\cdot\
\frac{\Big(U \bar S  \bar T^{2(m+1)} \bar S  \Big)_{\lambda\emptyset}
\Big(U \bar S  \bar T^{2(n+1)} \bar S  \Big)_{\lambda\emptyset}}
{ \underbrace{\Big(U \bar S  \bar T^{2} \bar S \, U ^{-1}\Big)_{\lambda\emptyset}
}_{ \Lambda'_\lambda=F_\lambda^{(1)}}}
\ee
\vspace{-0.4cm}
which actually implies that $U$ diagonalizes quadratic form with the matrix
$\bar S \bar T^{-2} \bar S  \bar T^{-2} \bar S$:\\
\be
\!\!\boxed{
\Lambda'_\lambda D_R \cdot
\Big((U^{\rm tr})^{-1}\, \bar S \bar T^{-2} \bar S  \bar T^{-2} \bar S \, U ^{-1}
\Big)_{\lambda \lambda'} = {Z^\lambda_R}\cdot \delta_{\lambda,\lambda'}
\ \  \Longleftrightarrow \ \
D_R\cdot\bar S \bar T^{-2} \bar S  \bar T^{-2} \bar S
= U^{\rm tr}\cdot{\rm diag}\Big( {\Lambda'}_\lambda^{-1} Z_R^\lambda\Big)\cdot U
}
\label{UZU}
\ee
Another possible implication is that $U$ provides
a prototype of the ("gauge invariant") arborescent vertex \cite{arborgauge}
\be
{\cal V} := \sum_\lambda \
\frac{\left.\bar S U^{-1}|\lambda\right>\otimes \left<\lambda|U \bar S \right.
\otimes \left<\lambda|U \bar S \right.}{ \Lambda'_\lambda}
\ee

{\bf 8.}
Like ${\bar S}$, matrix $U$ depends on $R$, we omitted the label $R$
to make the formulas readable. Universal ($R$-independent) are $\bar T$ and ${\cal B}$,
and just the first column $U_{Y\emptyset}=1$ of $U $.

\bigskip

{\bf Example} of $R=[1]$: From
\be
\bar S_{[1]} = \frac{1}{[N]}\left(\begin{array}{cc} 1 & \sqrt{[N+1][N-1]} \\ \sqrt{[N+1][N-1]} & -1
\end{array}\right), \ \ \
\bar T^2 = \left(\begin{array}{cc} 1 & 0 \\ 0 & A^2
\end{array}\right)
\ee
we get
\be
U_{[1]} = \left(\begin{array}{cc} 1 & \sqrt{[N+1][N-1]} \\ 1 &
\frac{A^2-(q^2-1+q^{-2})}{\{q\}^2\sqrt{[N+1][N-1]}}
\end{array}\right)
= \left(\begin{array}{cc} 1 & \sqrt{[N+1][N-1]} \\ 1 &
-\frac{1}{\sqrt{[N+1][N-1]}}+\frac{A}{\{q\}}\frac{[N]}{\sqrt{[N+1][N-1]}}
\end{array}\right), \ \ \
{\cal B} = \left(\begin{array}{ccc} 1 & 0  \\ -A^2 & A^2
\end{array}\right)
\ee
and
\vspace{-0.3cm}
\be
Z_{[1]}^\emptyset = 1, \ \ \ \ \ \ Z^{[1]}_{[1]} = \{Aq\}\{A/q\} = \{q\}^2[N+1][N-1]
\ee
To compare, before the $U$-"rotation", which converted it into {\it triangular} ${\cal B}$,
the original {\it symmetric} matrix was
\be
\bar S_{[1]} \bar T^2 \bar S_{[1]} = \frac{1}{[2] [N]}\left(\begin{array}{cc}
A^2\cdot (q^{-2}[N+1]+q^2[N-1])  & \ \ -A\{q^2\}  \sqrt{[N+1][N-1]}  \\ \\
-A\{q^2\}  \sqrt{[N+1][N-1]}  &  q^{2}[N+1]+q^{-2}[N-1]
\end{array}\right)
\ee
Note that {\bf the first line in $U$ is always the same as in $\bar S$},
i.e. consists of  the square roots of quantum dimensions of  the relevant representations
from $R\otimes \bar R$.
However, while $\bar S$ is finite,
$U$ is {\it singular} in the double-scaling limit when $q,A \longrightarrow 1$
and $N$ is fixed.
This is because $\bar T^2$ and thus $\bar S\bar T^2\bar S$ in this limit tend to a unit matrix,
which is preserved by any $U$-rotation, but does not satisfy (\ref{sumB}) --
thus, when approaching the limit, $U$ develops a singularity. $\bullet$

\newpage
\vspace{-0.3cm}

\bigskip
{\bf Example} of $R=[2]$: Likewise from
{\footnotesize
\be
\bar S_{[2]} =  \frac{[2]}{[N][N+1]}\left(\begin{array}{ccc}
1 & \sqrt{[N+1][N-1]} & \frac{[N]}{[2]}\sqrt{[N+3][N-1]}\\ \\
\sqrt{[N+1][N-1]} & \frac{[N+1]}{[2][N+2]}\Big([N+3][N-1]-1\Big)
&  -\frac{[N]}{[N+2]}\sqrt{[N+3][N+1]} \\ \\
\phantom.\frac{[N]}{[2]}\sqrt{[N+3][N-1]} & -\frac{[N]}{[N+2]}\sqrt{[N+3][N+1]}
&  \frac{[N]}{[N+2]}
\end{array}\right), \ \ \
\bar T^2 = \left(\begin{array}{ccc} 1 & 0 & 0  \\ 0 & A^2 & 0 \\ 0 & 0 & q^4A^4
\end{array}\right)
\nn
\ee
}

\noindent
it follows that

\bigskip

\centerline{
{\footnotesize
$
U_{[2]} =  \left(\begin{array}{ccc}
1 & \sqrt{[N+1][N-1]} & \frac{[N]}{[2]}\sqrt{[N+3][N-1]} \\ \\
1 & \frac{(q^6+q^4)A^4-(q^8+q^6-q^4+q^2+1)A^2+q^2}{q^4A^2\{q\}\{q^2\}[N+2]}
\sqrt{\frac{[N+1]}{[N-1]}} \!\!\!\!
&   \frac{A^2q^4-q^6+q^4-1}{q^3\{q\}\{q^2\}}\frac{[N]}{[N+2]} \sqrt{\frac{[N+3]}{[N-1]}} \\ \\
1 & \frac{A^2q^4-q^6+q^2-1}{q^3\{q\}^2[N+2]}\sqrt{\frac{[N+1]}{[N-1]}}
& \!\!\!\!\!\!\!\!\!\!\!
\frac{A^4q^{10}-A^2(q^{12}-q^8+q^6+q^4)+(q^{12}-q^{10}-q^8+2q^6-q^2+1)}{q^6\{q\}^3\{q^2\}
[N+2]\sqrt{[N+3][N-1]}}
\end{array}\right), \ \ \
{\cal B} = \left(\begin{array}{ccc} 1 & 0 & 0  \\ -A^2 & A^2 & 0 \\
q^2A^4 & -(q^2+q^4)A^4 & q^4A^4
\end{array}\right)
\nn
$
}}

\bigskip

\noindent
and
\vspace{-0.3cm}
\be
Z_{[2]}^\emptyset = 1, \ \ \ \ \ \ \ \ \
Z^{[1]}_{[2]} = [2]\{Aq^2\}\{A/q\} = \{q\}\{q^2\}[N+2][N-1],
\nn \\
Z^{[2]}_{[2]} = \{Aq^3\}\{Aq^2\}\{A\}\{A/q\} = \{q\}^4[N+3][N+2][N][N-1]
\ee
We see that, unlike $\bar S_R$ and $U_R$,
the matrices $\bar T$ and ${\cal B}$ for $R=[2]$ contain those for $R=[1]$
as sub-matrices -- this is a manifestation of their {\it universality}. $\bullet$

\bigskip

{\bf 9.} De facto, the KNTZ claim (\ref{evoF}) is that the switch
from a diagonal evolution matrix $\bar T^2$ to triangular
${\cal B}$, though looks like a complication,
actually reveals the hidden structure of the differential expansion for twist knots
and somehow trades the sophisticated Racah matrix $\bar S$ for a much simpler and
universal (representation-independent) ${\cal B}$.
As we explained, the reason for this can be that the actual evolution matrix
was not the simple diagonal $\bar T^2$,
but rather a sophisticated symmetric $\bar S\bar T^2\bar S$,
and then the switch to triangular ${\cal B}$ is indeed a simplification.
In this approach the crucial role is played by the switching matrix $U$,
and the central phenomenon is a drastic simplicity of the first line
in a peculiar matrix $U\bar T^2U^{-1}{\cal B}^{-1}$: for rectangular representations
its entries are just products of the differentials, and better understanding of
the phenomenon can help to explain the linear combinations of those, which emerge
in the non-rectangular case.

\bigskip

{\bf 10.}
The formula (\ref{evoF}) is very nice, still in this form it is not immediately
suitable for construction of Racah matrices with the help of (\ref{bS}).
Fortunately, this is easy to cure.
The simplest way to raise the matrix ${\cal B}$ to a power is to diagonalize it.
If ${\cal E}$ is a triangular matrix of right eigenvectors of ${\cal B}$,
\be
\sum_\mu {\cal B}_{\lambda\mu} {\cal E}_{\mu\nu} = {\cal E}_{\lambda\nu}\Lambda_\nu
\label{Edef}
\ee
then ${\cal B}^{m+1} = {\cal E} \Lambda^{m+1} {\cal E}^{-1}$
and  it follows from (\ref{evoF}) that
\be
\boxed{
f_{\lambda\mu} = {\cal E}_{\lambda\mu}\Lambda_\mu\sum_{\lambda'}{\cal E}^{-1}_{\mu\lambda'}
}
\label{fvsef}
\ee
Triangular ${\cal E}$ is defined modulo
right multiplication by a diagonal matrix
\be
{\cal E}_{\mu\nu} \longrightarrow {\cal E}_{\mu\nu}\cdot\xi_\nu
\label{gaugetr}
\ee
what can be used to convert diagonal elements of ${\cal E}$ into unities.
Expression (\ref{fvsef}) is invariant of this Abelian "gauge"  transformation
and for our purposes we are not obliged to make this additional conversion of ${\cal E}$.
When two or more eigenvalues $\Lambda_\mu$ coincide, the ambiguity in  ${\cal E}$ increases.
One can choose the corresponding block to be a unit matrix.

\bigskip

{\bf 11.}
Alternative expression for Racah matrix $\bar S$ is provided by a product
of  (\ref{UZU}) and (\ref{BviaU}):
\be
D_R \cdot \bar S  =
U^{\rm tr}\cdot{\rm diag}\Big( {\Lambda'}_\lambda^{-1} Z_R^\lambda\Big)
\cdot {\cal B} \cdot U \cdot   \bar T^2
\ \ \ \ \ \ \ \ \ {\rm i.e.} \ \ \ \ \ \
 \bar S_{\mu\nu}  = \frac{1}{D_R}\cdot
\sum_{\lambda,\rho}  {\Lambda'}_\lambda^{-1} Z_R^\lambda \cdot U_{\lambda\mu}
{\cal B}_{\lambda\rho}U_{\rho\nu}\Lambda_\nu
\ee
\vspace{-0.3cm}
It, however, requires an explicit knowledge of the matrix $U$ -- in addition
to the generally known $Z$, $\Lambda$ and ${\cal B}$.

\newpage

{\bf 12. }
Generalization to {\bf 4-graded hyperpolynomials} \cite{GGS,diffarth,NawOb},
i.e. the $\beta$-deformation \cite{betadefo} of (\ref{difexpan}),
is straightforward and  follows
the recipe of \cite{evo,diffarth} and \cite{KM17fe,KM17tw}: \\

(a) in differentials $\{Aq^{i-j}\}$
were $i$ and $j$ are associated with the arms and legs in the Young diagram,
the positive and negative powers of $q$
become the powers of $t=q^\beta$ instead of $q$: $\{A q^i/t^j\}$,\\

(b) in $Z$-factors (\ref{Zfactor}) quantum dimensions are substituted by Macdonald ones,\\

(c) the two constituents of $Z$-factors acquire opposite powers $\sigma^{\pm 1}$ of
the forth grading parameter, which appears nowhere else in the formulas, \\

(d) skew Schur functions in $F$'s are substituted by skew Macdonalds and\\

(e) dimensions $\Lambda_\lambda$ are $\beta$-deformed,
$\Lambda_\lambda \longrightarrow \tilde\Lambda_\lambda$. \\

\noindent
Minor new comments are needed only at the points (d) and (e).
In (d) one should choose the proper version of Schur formula to deform.
Suggested in \cite{KNTZ} was a complicated version, based on the possibility to re-express
skew Schur functions at the $zero$-locus through {\it shifted} Schur functions
\cite{ShiSchur}--
so that the $\beta$-deformation involves shifted (interpolating) Macdonald polynomials
\cite{ShiMac}.
Remarkably, this works, but there is a much simpler option:
transposed skew Schurs in (\ref{Bels}) can be substituted by those of negative times --
and then they can be substituted by skew Macdonalds (at Macdonald level transposition
differs from time-inversion by somewhat complicated additional factors -- and it is best
to find the formulation where they do not show up).

The point (e) is more tricky.
The   hook formula (\ref{Lambda}) is actually equivalent to
the $q$-power of {\it regularized}  Casimir \cite{MMuniv}:
$\varkappa_{(\lambda,\lambda)} - \varkappa_{[\lambda_1^N]}$.
where $\varkappa_R = 2\sum_{(i,j)\in R} (i-j)$
and $\lambda_1$ is the longest line in $\lambda$.
The $N^2$ contribution is eliminated by taking the difference,
which is equal to  $N|\lambda| + c_\lambda$, and $c_\lambda$ is best
expressed in terms of hooks.
As known since \cite{DMMSS}, the $\beta$-deformation splits $\varkappa_R = 2(\nu_{R^\vee}-\nu_R)$
with $\nu_{R^\vee} = \sum_{(i,j)\in R}(i-1)=\sum (i-1)R^\vee_i$
and $\nu_R = \sum_{(i,j)\in R}(j-1)=\sum (j-1)R_j$.
The actual formula for the $\beta$-deformed eigenvalue is
\be
\tilde\Lambda_\lambda = q^{2(\nu_{(\lambda,\lambda)^\vee}-\nu_{[N^{\lambda_1}]})}
\cdot t^{-2(\nu_{(\lambda,\lambda)}-\nu_{[\lambda_1^N]})}
\cdot \left(\frac{A^2}{qt^{2N-1}}\right)^{|\lambda|}
\label{tildeLambda}
\ee
and it is actually independent of $N$.
For symmetric and antisymmetric representations $\lambda=[r]$ and $\lambda=[1^s]$
expressions (\ref{tildeLambda}) reproduces the $A^2\longrightarrow A^2q/t$ prescription
of \cite{evo},
but for generic rectangular representations $\lambda=[r^s]$ it provides explicit,
but less trivial expressions.

Putting everything together, we obtain the following generalization of eq.(40)
of \cite{KM17fe} for the rectangularly-colored
4-graded hyperpolynomial from the figure-eight to arbitrary twist knots:
\be
\boxed{
{\cal P}^{(m,n)}_{[r^s]}(A,q,t,\sigma) \
= \sum_{\lambda\subset [r^s]} {\cal M}^\vee_{\lambda^\vee}\{p^*(r)\}{\cal M}_{\lambda}\{p^*(s)\}
\cdot  \prod_{(i,j)\in\lambda} \left\{\frac{Aq^{r+i}}{\sigma t^j}\right\}
\left\{\frac{A\sigma q^i}{t^{s+j}}\right\}  \cdot
\tilde{\cal F}_\lambda^{(m,n)}
}
\label{hyperp}
\ee
with the factorized
\be
\boxed{
\tilde{\cal F}_\lambda^{(m,n)} =
\frac{\sum_{\mu\subset\lambda} (\tilde{\cal B}^{m+1})_{\lambda\mu}
\sum_{\nu\subset\lambda} (\tilde{\cal B}^{n+1})_{\lambda\nu}}
{\tilde {\cal B}_{\lambda\emptyset}}
}
\ee
and the $\beta$-deformed KNTZ-like triangular evolution matrix
\be
\boxed{
\tilde{\cal B}_{\lambda\mu}
= \tilde\Lambda_\lambda \cdot \frac{{\cal M}_{\lambda/\mu}\{-\tilde p^\circ\}\cdot
{\cal M}_{\mu}\{\tilde p^\circ\}}{{\cal M}_\lambda\{\tilde p^\circ\}}
}
\label{tilB}
\ee
and the $\beta$-deformed zero-locus
\vspace{-0.3cm}
\be
\tilde p^0_k = \frac{\{q\}^k}{\{t^k\}}
\ee
Here ${\cal M}$ denotes Macdonald polynomials \cite{Mac},
${\cal M}^\vee(q,t) := {\cal M}(t^{-1},q^{-1})$, and we use the standard
condensed notation $\{x\}:=x-x^{-1}$.

\newpage

For example, the matrix ${\cal B}$ for representation $R=[2]$ from the second example
in sec.8 gets $\beta$-deformed in the following way:
\be
{\cal B}_{[2]} = \left(\begin{array}{ccc} 1 & 0 & 0 \\ \\ -A^2 & A^2 & 0 \\ \\
q^2A^4 & -(q^2+q^4)A^4 & q^4A^4
\end{array}\right) \ \ \ \stackrel{t\neq q}{\longrightarrow} \  \ \
\tilde{\cal B}_{[2]} = \left(\begin{array}{ccc} 1 & 0 & 0  \\ \\
-\frac{qA^2}{t} & \frac{qA^2}{t} & 0 \\ \\
\frac{q^4A^4}{t^2} & -\frac{q^4(q^2+1)A^4}{t^2} & \frac{q^6A^4}{t^2}
\end{array}\right)
\ee
\\

\noindent
Similarly, for representation $R=[2,2]$ \\
\be
\!\!\!\!\!
\tilde{\cal B}_{[2,2]} = \left(\begin{array}{cccccc}
\emptyset & [1] & [1,1] & [2] & [2,1] & [2,2] \\ \\
\hline \\
1 & 0 & 0 & 0 & 0 & 0  \\ \\
-\frac{qA^2}{t} & \frac{qA^2}{t} & 0 & 0 & 0 & 0\\ \\
\frac{q^2A^4}{t^4} & -\frac{q^2(t^2+1)A^4}{t^6} & \frac{q^2A^4}{t^6} & 0 & 0 & 0 \\ \\
\frac{q^4A^4}{t^2} & -\frac{q^4(q^2+1)A^4}{t^2} & 0 & \frac{q^6A^4}{t^2} & 0 & 0 \\ \\
-\frac{q^5A^6}{t^5} & \frac{q^5(q^2t^2+t^2+1)A^6}{t^7} & -\frac{(q^4t^2-1)q^5A^6}{t^7(q^2t^2-1)}
& -\frac{(q^2t^4-1)q^7A^6}{t^7(q^2t^2-1)} & \frac{q^7A^6}{t^7} & 0 \\ \\
\frac{q^8A^8}{t^8} & -\frac{(q^2+1)(t^2+1)q^8A^8}{t^{10}} &
\frac{(q^2+1)(q^4t^2-1)q^8A^8}{t^{10}(q^2t^2-1)}
& \frac{(t^2+1)(q^2t^4-1)q^{10}A^8}{t^{12}(q^2t^2-1)}
& -\frac{(q^2+1)(t^2+1)q^{10}A^8}{t^{12}} & \frac{q^{12}A^8}{t^{12}}
\end{array}\right) \,
\ee

\bigskip

\noindent
As clear from this example, the  KNTZ matrix has non-polynomial entries.
Still for all rectangular $R=[r^s]$ we get hyper{\it polynomials},
moreover they become {\it positive}
in the DGR variables \cite{DGR} (boldfaced): $A^2=-{\bf a}^2{\bf t}$, $q=-{\bf qt}$, $t={\bf q}$,
i.e. can actually pretend to be the {\it super}polynomials
-- for all twist, and, actually, double-braid knots \cite{KNTZ}.

The entries of the first column in $\tilde{\cal B}$
control the trefoil, which is a member of the twist family with $m=1$,
since (\ref{sumB}) survives after the $\beta$-deformation:
\be
\sum_\mu \tilde{\cal B}_{\lambda\mu} = \delta_{\lambda,\empty}
\ \Longrightarrow \
\sum_\mu (\tilde{\cal B}^2)_{\lambda\mu} = \tilde{\cal B}_{\lambda\emptyset}
\ee
They are always polynomial and given by a simple hook-type formula  \cite{KM17fe}:
\be
\tilde{\cal B}_{\lambda\emptyset}
= \tilde\Lambda_\lambda \cdot \frac{{\cal M}_{\lambda }\{-\tilde p^\circ\}}
{{\cal M}_\lambda\{\tilde p^\circ\}}
= \left(-\frac{A^2q}{t}\right)^{|\lambda|}\cdot\prod_{(i,j)\in\lambda} q^{2(R^\vee_i-i)}
t^{-2(R_j-j)}
\ee


{\bf 13. Conclusion and open problems.} The two conjectures (\ref{difexpan}) and (\ref{evoF})
together with explicit formulas (\ref{Zfactor}) and (\ref{Bels})
provide a {\bf closed explicit expression for arbitrary rectangularly-colored
HOMFLY-PT polynomials for the double-twist knots},
while  (\ref{hyperp})--(\ref{tilB}) do the same for their {\bf 4-graded
hyperpolynomial deformation}. \\

{\bf a)} Still both conjectures need a proof, at least in the Reshetikhin-Turaev formalism
\cite{RT}, and finally -- at the level of Chern-Simons theory \cite{CS}.
Of most interest would be the (still lacking) explanation,
why the coefficients of the differential expansion factorize in (\ref{difexpan}).\\

{\bf b)} Generalization to other knots is desirable, especially to
the next-in-the-line family of pretzel knots \cite{pretzel}.\\

{\bf c)} A better understanding of the  Rosso-Jones-like \cite{RJ,DMMSS} formula (\ref{RJan})
would be useful,
in particular the mysterious conspiracy between the
somewhat  strange projector (\ref{UY0}) onto the vector $(1,\ldots,1)$,
and the factorization of $Z$-factors (\ref{ZviaU}), provided by the $U$-rotation of
the vector $\langle \emptyset|\bar S\bar T^2\bar S\bar T^{-2}\bar S$, \ which by itself has
quite complicated components.

\newpage

{\bf d)} As explained in sec.11, the generalization to {\it hyper}polynomials is straightforward,
once the three-level structure
\be
\text{ factorized differential expansion (\ref{difexpan})}
\ \longrightarrow \ \text{matrix evolution   (\ref{evoF})}
\ \longrightarrow \  \text{skew characters formula (\ref{Bels})}
\nn
\ee
is revealed.
Since rectangularly-colored hyperpolynomials for all twist knots are  positive, as expected,
this highly extends the set of rectangularly-colored {\it super}polynomials,
which  so far were rarely available beyond torus knots.
Still it is interesting to understand the reasons and the proof of polynomiality
and positivity -- both are far from obvious in the present approach.


\bigskip

In the case of Racah matrices the formulas (\ref{bS}), (\ref{fvsef}) and (\ref{SfrobS})
are currently less explicit:
the procedure involves solving two linear systems --
(\ref{Edef}) for ${\cal E}$ and then (\ref{SfrobS}) for $S$. \\

{\bf e)}
Still, given the simple form (\ref{Bels}) of ${\cal B}$,
one can  hope to get more explicit expressions for
the {\it universal} triangular matrices
${\cal E}$ and ${\cal E}^{-1}$ of the right and the left eigenvectors
and then -- for $\bar S$ and $S$. \\

{\bf f)}  Orthogonality of ${\bar S}$ is far from obvious --
and serves as an indirect and non-trivial check of the three-step conjecture in {\bf d)}.
Orthogonality of $S$ is straightforward, once it is defined as a diagonalizing
matrix from (\ref{SfrobS}) and properly normalized, i.e. Abelian gauge freedom like
(\ref{gaugetr}) is used to make it orthogonal. \\

{\bf g)} Developing the considerations of \cite{hypergeom},
one can look for the hypergeometric interpretations of Racah matrices --
at least for rectangular representations, where they are unambiguously defined.\\

{\bf h)} One can wonder what happens to relation between (\ref{RJan}) and (\ref{HRdb})
in the case of hyperpolynomials -- and what is the hyper-analogue of ${\bar S}$
which one could define in this way from explicit expressions
(\ref{hyperp})--(\ref{tilB}).
It is already interesting  to see what happens to orthogonality of this matrix.

\bigskip

The most puzzling is generalization from rectangular to arbitrary representations.
For rectangular $R=[r^s]$ all the representations in $R\otimes \bar R$ are
in one-to-one correspondence with the sub-diagrams $\lambda$ of $R$,
what is heavily used in all the above formulas.
For generic $R$, this relation is no longer one-to-one -- one can say that
some $\lambda\subset R$ enter with non-trivial multiplicities $c_\lambda$.
As explained and illustrated in details in \cite{Mnonrect}
this is not a very big problem at the level of colored knot polynomials,
still the general answer is not yet available. \\

{\bf i)} One can now look for a generalization of the matrix evolution formulas (\ref{evoF})
for generic $R$ and for explicit expression for the corresponding HOMFLY-PT,
see \cite{M19nonrect} for a recent progress in this direction.
An important issue here is construction of the $U$-matrix and further
decomposition of the  $Z_R^X$ with the help of (\ref{ZviaU}). \\

{\bf j)} One can expect that associated 4-graded hyperpolynomials will fail to be
positive -- and this would provide a whole family of non-trivial answers
beyond the torus-knot set, what can help to better understand the
problem of colored superpolynomials \cite{DGR,NawOb},
as well as the relation to the evolution properties of Khovanov-Rozansky polynomials,
studied in \cite{Anokhevo}.

\bigskip

A more severe problem in non-rectangular case is that
arborescent formula (\ref{HRdb})  no longer distinguishes between the
elements in entire $c_\mu\times c_\nu$ blocks in the matrix $\bar S$,
thus one can not extract it directly from explicit expression for the double-twist family.
\\

{\bf k)} One can wonder if $\bar S$ can still be found on this path,
as it was done with the help of non-linear and therefore difficult unitarity requirement
in the simplest cases of $R=[2,1]$ and $R=[3,1]$ in \cite{Mnonrect}.
The most interesting, of course, is the  case of $R=[4,2]$ -- the first one
where multiplicities are not separated by symmetries.
Additional help here could be provided by a solution of the problem {\bf b)}.

\section*{Acknowledgements}

My work is partly supported by the grant of the
Foundation for the Advancement of Theoretical Physics BASIS,
by RFBR grant 19-02-00815
and by the joint grants 17-51-50051-YaF, 18-51-05015-Arm,
18-51-45010-Ind,  RFBR-GFEN 19-51-53014.
I also acknowledge the hospitality of KITP and
partial support by the National Science Foundation
under Grant No. NSF PHY-1748958
at a certain stage of this project.

\end{document}